\documentstyle[prl,aps,psfig,twocolumn]{revtex} 
\newcommand{\avg}[1]{\langle {#1} \rangle}
\def\be{\begin{equation}}
\def\ee{\end{equation}}
\begin{document}
\twocolumn[\hsize\textwidth\columnwidth\hsize\csname@twocolumnfalse\endcsname
\title{Quantitative description and modeling of real networks.}
\author{Andrea Capocci$^1$, Guido Caldarelli$^2$, Paolo De Los Rios$^3$.}
\address{$^1$D\'epartement de Physique,
Universit\'e de Fribourg, CH-1700 Fribourg, Switzerland.}
\address{$^2$INFM UdR di ROMA1 Dip. Fisica, Universit\`a di Roma
``La Sapienza'' P.le A. Moro 2 00185 Roma, Italy.}
\address{$^3$Institut de Physique Th\'eorique, Universit\'e de Lausanne, 
1015 Lausanne, Switzerland. INFM UdR Torino Politecnico, Dip. Fisica, Politecnico di Torino, Corso duca degli Abruzzi 24, Torino, Italy}
\date{\today}
\maketitle
\begin{abstract}
In this letter we present data analysis and modeling of two particular
cases of study in the field of growing networks. We analyze WWW data set and
authorship collaboration networks in order to check the presence of correlation 
in the data. The results are reproduced with a pretty good agreement 
through a suitable modification of
the standard AB model of network growth.
In particular, intrinsic relevance of sites plays a role in determining
the future degree of the vertex. 
\end{abstract}
]

\narrowtext

The fractal properties of social networks have
been largely investigated by the statistical
mechanics community in recent times.
Many quantities have been recognized as ``signatures''
of complexity in such networks.
In particular, the probability distribution 
of the degree of the nodes in a social network 
displays an algebraic decay in several different 
realizations, including the
Internet, the WWW, the movie actors network and the
science collaboration network
\cite{faloutsos,caldarelli,barabasi,newman1}.
Since then, many models have been developed in order to
reproduce this particular feature of real networks
\cite{barabasi,huberman}.

Recent studies have provided a more detailed picture
of the connectivity in social networks.
In the Internet Autonomous Systems (IAS) network, the relation
between the degree of a node $k$ and the average degree
of its neighbors $k_{nn}(k)$ has been measured, 
showing a decaying behavior of $k_{nn}(k)$ for 
large $k$; such property is connected to a hierarchical 
structure of the growth process \cite{goh,vespignani}.

Using a slightly different formalism, it has been shown 
that a taxonomy of social networks can be made according
to the correlation between the degrees of directly 
connected nodes \cite{newman2}.
In networks displaying ``assortative (disassortative) 
mixing'', the correlation is positive (negative),
which corresponds to an increasing (a decreasing) behavior 
of $k_{nn}(k)$.

Moreover, a growing number of researches deals with the
clustering properties of social networks, that is,
the presence and the abundance of groups of nodes having
a strong internal connectivity.
To study the clustering properties, we lack a unique
physical quantity: directed and undirected graphs,
indeed, require different approaches.
In the undirected case, the clustering coefficient,
i.e. the average fraction of neighbors of a node
that are also directly connected one to each other,
is usually measured.

Recent surveys on IAS \cite{goh,vespignani} have
measured the clustering coefficient $c_{k}$ around 
nodes of degree $k$.
These empirical studies show a decaying behavior of $c_k$
with respect to $k$, as in the case of $k_{nn}(k)$.

The same quantities can be measured in directed graphs, 
though the generalization to this case may be somewhat 
arbitrary.
In principle, one could consider the in-going or the outgoing
links in finding the neighbors of a node to measure their 
degree and their mutual connectivity.
This way, the number of directed links within a group
of nodes may be greater than the number of pairs of node,
thus leading to clustering coefficient greater than one.

Another and simpler way to generalize such a method to directed graphs 
is to take one-way links as bidirectional ones, and to consider the 
resulting undirected graphs, where the traditional definitions
apply.
We neglect that some pair of nodes may be actually mutually 
linked, and we replace this two directed links with a single 
undirected one. 
In the WWW database we used \cite{wwwdata}, for example, about one fifth 
of the links are reciprocal.
We have adopted this technique to measure both $k_{nn}(k)$
and $c_{k}$, finding qualitatively the same results as
in the IAS (undirected) case studied in \cite{goh,vespignani}.
Both quantities behave as a power law for large $k$, with decay 
exponents close to the ones measured in \cite{goh,vespignani}.

\begin{figure}
\centerline{\psfig{file=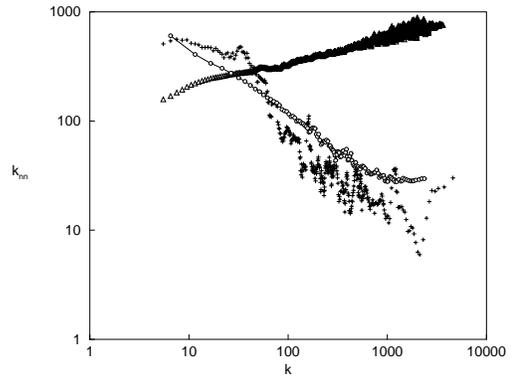,width=6.5cm}}
\caption{
Average degree $k_{nn}(k)$ of nearest neighbors
of a node with degree $k$, as a function of $k$. 
Triangles refer to the actor collaboration network, 
plus symbols refer to the WWW empirical survey
(10-points averaged), circles to simulations
of our model.}
\label{fig1}
\end{figure}

\begin{figure}
\centerline{\psfig{file=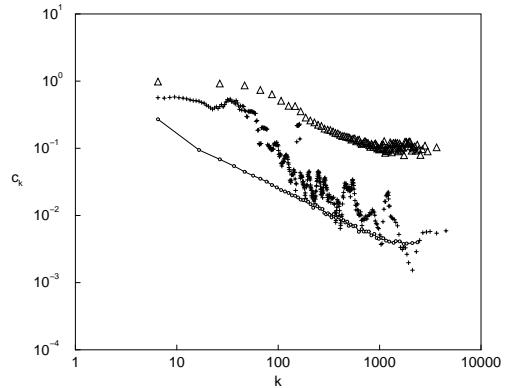,width=6.5cm}}
\caption{
Clustering coefficient around a node of degree 
$k$ as a function of $k$.
Circles refer to the actor collaboration network, 
plus symbols refer to the WWW empirical survey
(10-points averaged) and squares to the simulation
of our model.
}
\label{fig2}
\end{figure}

Standard noise reducing data analysis techniques show that 
$k_{nn}(k) \sim  k^{-0.76}$ for large $k$, as shown in Figure \ref{fig1}, 
and $c_{k} \sim  k^{-1.03}$, see Figure \ref{fig2}.

This behavior is in qualitative agreement with the
power laws found in the IAS case \cite{vespignani}, 
though the exponents are slightly different; By their
measurements, which are affected by a weaker noise,
$k_{nn}(k) \sim  k^{-0.5}$ and
$c_{k} \sim  k^{-0.75}$.

This phenomenon, however, is not ubiquitous. 
Indeed, it is observed in networks where the 
decision of connecting a pair of nodes only 
depends on one of the connecting node.
We claim that the distinction between ``assortative''
and ``disassortative'' mixing, as introduced in
\cite{newman2}, relies on this particular property 
of the microscopic growth mechanism.

In the WWW and in the IAS networks, the link growth 
mechanism is strictly local, lacking any outer supervision. 
In this case, each node is free to choose highly relevant
neighbors. 

On the other hand, networks with assortative mixing are 
often examples of networks where a single node has no 
power to choose its neighbors. 
E.g., in the actors collaboration networks film directors
decide the link structure and the nodes, the actors, 
have no power to direct their connectivity.
Due to economical constraints, expensive celebrities
are often balanced by less relevant actor, biasing
the connectivity correlation.
A qualitative difference is found in $k_{nn}(k)$ and 
in $c_k$ for this particular network.
As shown in Figure \ref{fig1}, $k_{nn}(k)$ grows with
$k$, in contrast with the decay observed in the IAS
and in the WWW.
This confirms what has been empirically found in \cite{newman2} 
where the WWW appears among the network with disassortative
mixing, whereas the actor collaboration network has
assortative mixing.
$c_k$, however, displays a decreasing behavior, though
it does not seem to follow a power law as in the the IAS
and WWW cases.

Our claim is reinforced by other cases of assortative mixing, 
studied in \cite{newman2}, like the scientific collaboration 
networks. 
Indeed, to establish a scientific collaboration the agreement 
of both scientists is needed, and a single node of such 
networks is not free of choosing its neighbors.

To check if our hypothesis is true, we introduce a growing
undirected network model.
Sites are added at a discrete pace, and each site has an
intrinsic ``relevance'', which is a random variable drawn from 
a uniform distribution in the range $[0,1]$.

In our interpretation, a link is a relevance attribution 
to the pointed node, in the spirit of \cite{brin,lifantsev,kleinberg}.
In the WWW, for example, a relevant web page rarely points to a
non relevant one, suggesting a relevance-driven connectivity 
concentration.
To implement such a policy, in our model a node added at time 
$t$ with a relevance $r_t$ can be connected only to nodes 
having a relevance higher than $r_t$, with linear preferential 
attachment: the probability of acquiring a new link is 
proportional to the actual degree.

This implies that an existing node $i$ with a relevance $r_i$ and
degree $k_i$ has a probability
$p_i = \Theta(r_s - r_t)\frac{k_i}{\sum_{s=1}^t k_s \Theta(r_s - r_t) }$
of acquiring a new link, where $\Theta(x)=1$ for $x>0$ and $\Theta(x)=0$ 
otherwise.
Finally, we assume that a newly added node is connected to $m$ existing nodes
according to the described rule.

Let us call $k_i(t)$ the degree at time $t$ of the node
$i$ introduced at time $i$, whose ``relevance'' is $r_i$.
At each time step, there is a probability $r_{i}$ that the
newly introduced node has a relevance $r_t<r_i$, since
$r_t$ is drawn from a uniform distribution between $0$ and $1$.
Then, the probability of increasing by $1$ the degree $k_i(r_i,t)$
is approximately given by
\begin{equation}
\avg{p_i}_{r_t} \simeq \frac{r_i k_i(t)}{\avg{\sum_{s:r_s>r_t}^{1,t-1} k_s(t)}_{r_t}},
\end{equation}
where $\avg{.}_{r_t}$ denotes the average over all the realizations
of $r_t$.
In the following, we will neglect the explicit time dependence
whenever unnecessary.
We can write a rate equation for the degree, following the
reasoning made in \cite{bianconi}:
\begin{equation}
\dot{k_{i}} = \frac{m r_i k_i(t)}{\avg{\sum_{s:r_s>r_t}^{1,t-1} k_s(t)}_{r_t}}.
\end{equation}
To evaluate the denominator in the r.h.s. of the equation above, we
have to compute
\begin{equation} \label{denominator}
\avg{\sum_{s:r_s>r_t}^{1,t-1} k_s(t)}_{r_t} = \int_0^{r_i} dr_t A(r_t,t),
\end{equation}
where we defined
\begin{equation} 
A(r_t,t) = \sum_{s:r_s>r_t}^{1,t-1} k_s(t).
\end{equation}

$A(r_t,t)$ is a decreasing function of $r_t$.
If $r_t=0$, $A(r_t,t)$ is the sum of the degree of all the nodes
at time $t$, i.e. $A(0,t)=2mt$; on the other hand, if $r_t=1$
the sum in the definition of $A(r_t,t)$ does not contain any
term; therefore, $A(1,t)=0$.
Thus, we assume 
\begin{equation} \label{ansatz}
A(r_t,t)=2mt(1-r_t^{\alpha(t)})
\end{equation} 
as the general functional form of $A(r_t,t)$.
By this ansatz, we can compute the r.h.s. of eq.(\ref{denominator}),
\begin{equation}
\int_0^{r_i} dr_t A(r_t,t) = 2mt r_i(1-\frac{r_i^{\alpha(t)}}{1+\alpha(t)}).
\end{equation}
Let us assume that 
\begin{equation}
\alpha = \lim_ {t \to \infty} \alpha(t).
\end{equation}
In this case, we can define
\begin{equation}
C(r) = 1-\frac{r^{\alpha}}{1+\alpha}.
\end{equation}
Therefore, for large $t$ the rate equation takes the form
$\dot{k_i} = \frac{r_i k_i}{2tC(r_i)}$
which admits the solution
\begin{equation} \label{tevol}
k_i(t) = m \left( \frac{t}{i} \right) ^{\frac{r_i}{2C(r_i)}}
\end{equation} 
for the time evolution of the degree, following the same 
reasoning as in \cite{barabasi}.

Let us now call $K(r,t)dr$ the the sum of the degrees of
the nodes with relevance between $r$ and $r+dr$, at time $t$.
At each time step, $dr$ nodes on average are introduced with 
such a relevance. 
Eq. (\ref{tevol}) gives us the degree acquired by each of
these nodes.
To obtain $K(r,t)$ we have to sum over all time steps
from $1$ to $t$, and we get
\begin{equation}
dr K(r,t) = dr \sum_{s=1}^t k_s(t).
\end{equation}
If $s$ is continuous, the sum becomes an integral 
\begin{equation} \label{density}
dr K(r,t) = mdr \int_0^t ds \left(\frac{t}{s}\right)^\frac{r}{2C(r_{s})}.
\end{equation}
which implies
\begin{equation}
K(r,t) = \frac{mt}{1-\frac{r}{2C(r_s)}}.
\end{equation}

\begin{figure}
\centerline{\psfig{file=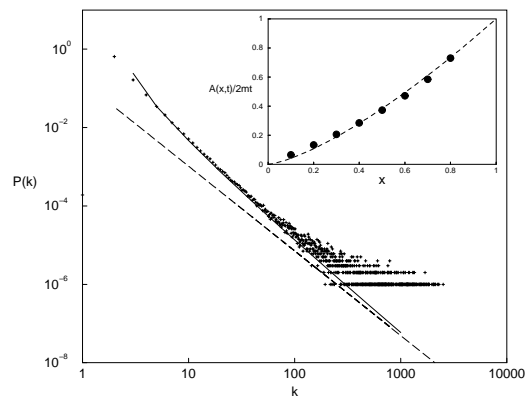,width=6.5cm}}
\caption{
Degree PDF in our network model made of $10^4$ nodes, 
with $m=2$.
Plus symbols refer to numerical simulation.
The solid line is obtained by plotting eq. \ref{pdf}.
The dashed line is proportional to $k^{-2.16}$.
Inset: The function $\frac{A(x,t)}{2mt}$ plotted for $t=10^4$ 
and $m=2$.
The dashed line represents $x^{1.38}$, displayed here 
to check the validity of our ansatz.
}
\label{fig3}
\end{figure}

We can estimate $\alpha$ by integrating $K(r,t)$ over all $r$,
thus obtaining the total sum of the nodes' degree:
\begin{equation}
\int_0^1 dr K(r,t) = 2mt.
\end{equation}
Therefore, using the expression of $K(r,t)$ we can write
the following equation for $\alpha$,
\begin{equation}
\int_0^1 dr \frac{1+\alpha -r^{\alpha}}{(2-r)(1+\alpha)-2r^{\alpha}}=1,
\end{equation}
This equation can be solved numerically, and yields 
$\alpha=1.3837$.
The hypothesis made in equation (\ref{ansatz}) is verified
in simulations of the model, as shown in Figure \ref{fig3} 

Following \cite{bianconi}, we compute the statistical
distribution of the degree $P(k)$ by its time evolution.
The probability $P(k_i(t)>k)$ that a randomly chosen node 
$i$ has a degree higher than $k$ at time $t$ is equal to 
the probability that the node has been introduced in the 
network at a time
$i < t (k/m)^{-\frac{2C(r_i)}{r_i}}$,
as one may verify by solving the time evolution of $k_i$
with respect to $i$.
Since nodes are added at a uniform pace, we have
\begin{equation}
P(k_i(t)>k) = (k/m)^{-\frac{2C(r_i)}{r_i}}.
\end{equation}
By definition, the probability distribution
of the degree of nodes with relevance $r$ is
$P(k,r_i) = -\frac{d}{dk}P(k_i(t)>k)$.

The total degree distribution, regardless
the relevance of the node, is obtained by averaging $P(k,r_i)$
with respect to the uniform distribution of $r_i$:
\begin{equation}
P(k) = - \int_0^1 dr_i \frac{d}{dk} P(k_i(t)>k)
\end{equation}
After replacing the kernel of this integral by
its explicit expression obtained above, we get
\begin{equation} \label{pdf}
P(k) = 
\frac{1}{k} \int_0^1 dr \left(\frac{k}{2}\right)^{-B(r)}B(r),
\end{equation}
where $B(r)= \frac{2}{r}(1-\frac{r^{\alpha}}{1+\alpha})$.

We can estimate the power-law exponent of the degree distribution $P(k)$
finding upper and lower bounds for its integral expression.
Indeed, we find that 
\begin{equation}
F(r) = e^{-2 \ln(k/2) B(r)} B(r) 
\end{equation}
is such that the integrand is monotonically growing.
Therefore it is easily seen that
\begin{equation}
P(k) < \frac{1}{k} e^{-2 \ln(k/2) B(1)} B(1) \sim k^{-\frac{3\alpha+1}{\alpha+1}}
\end{equation}
As for the lower bound, we first observe that the integrand is monotonically
increasing, with positive second derivative. So, 
\begin{equation}
F_1(r) = F(1) - F'(1) (1-r)
\end{equation}
is such that $F_1(r) < F(r)$ for $0 \le r \le 1$. 
If we then extend the integral from $r_1$ ($F_1(r_1) = 0$) 
to $1$, we surely find an underestimate for $P(k)$.
In particular we find
\begin{equation}
P(k) > k^{-\frac{3\alpha+1}{\alpha+1}} \frac{1}{\frac{2 \alpha}{\alpha+1} \ln(k/2) -1}
\end{equation}
The asymptotic behavior of $P(k)$ is 
therefore $k^{-\frac{3\alpha+1}{\alpha+1}}$
with at most logarithmic corrections.

We numerically checked that $P(k)$ is a power law 
with a rather weak correction that slows down 
the decay, as displayed in Figure \ref{fig3}.
Neglecting the correction, the best approximating 
exponent of the PDF is about $-2.16$, which confirms
the above computation. 
Indeed, we have $\frac{3\alpha+1}{\alpha+1} = 2.16$.

This value, moreover, is close to the exponents one
measures in real networks, which lie in the
range $2-2.4$.

In the simulation of the model, $k_{nn}(k)$ and
$c_k$ have also been numerically investigated. 
Unfortunately, we could not find an analytical description of
these two quantities.
As required by real data, $k_{nn}(k)$
and $c_k$ decay algebraically with respect to $k$.
For the nearest neighbors degree, we approximately 
measured
$k_{nn}(k) \simeq k^{-0.57}$,
as shown in Figure \ref{fig1}.
The value of the exponent agrees with the 
measurement reported in \cite{vespignani,goh}, 
which yields $k_{nn}(k) \simeq k^{-\nu_k}$ with
$\nu_k=0.5 \pm 0.1$.

As of the clustering coefficient $c_k$, simulations
reported in Figure\ref{fig2} show that $c_k \simeq k^{-0.72}$.
The same relation, measured by \cite{vespignani,goh}, 
in the IAS networks case, reads  $c_k \simeq k^{-\omega}$
with $\omega = 0.75 \pm 0.03$.

The qualitative behavior of these quantities is reproduced
in our extremely simple model.
As a comparison, let us recall that, without an intrinsic relevance,
a simple growing network model with preferential
attachment shows no correlation between the degrees
of two linked nodes. In addition, in this models
the clustering coefficient around a node does not
depend on the the degree of the node \cite{vespignani,newman2}.

An improvement in approximating real data could be achieved by 
adding other microscopic interactions to the dynamics
of our toy model, such as rewiring and elimination
and links, or by merging nodes, as already done in
former works \cite{albert,dorogotsev,capocci}
in the search for a better approximation of the scale free
degree distribution.

We believe that our analysis has pointed out some key
structural features of social networks, by the observation
of the correlation and the clustering of the connectivity
in networks.
In particular, the non trivial behavior of the nearest-neighbor
average degree and of the connectivity coefficient
have been measured in some real examples.
We also provided a toy model is a growing networks with 
preferential attachment, where nodes only connect to more 
relevant ones.
We have shown numerically and analytically, as far as we could, 
that our model reproduces qualitatively the statistical
properties of real networks, including the correlations
in the connectivity.
We believe that this approach suggests new empirical measurements
to be carried out on real networks, as well as needs new 
analytical steps further in the comprehension of this
complex systems.

The authors wish to thank M.E.J. Newman, A.-L. Barabasi, 
R. Albert, H. Jeong for providing free-access data about networks. 
They acknowledge support form EC-Fet Open project COSIN IST-2001-33555, and the OFES-Bern(CH).
G.C. and P.D.L.R. finally wish to thank the Herbette Foundation for 
financial support.

\end{document}